\documentclass[%
groupedaddress,
amsmath,amssymb,
aps,
twocolumn,
prl,
floatfix,
notitlepage
]{revtex4-1}

\usepackage{graphicx}% Include figure files
\usepackage{dcolumn}% Align table columns on decimal point
\usepackage{bm}% bold math
\usepackage{subfigure}%subfigure enviroment
\makeatletter
\renewcommand{\thesubfigure}{\alph{subfigure}} %(a) -> a
\renewcommand{\@thesubfigure}{(\thesubfigure)\hskip\subfiglabelskip} %a -> a)
\makeatother
\usepackage[colorlinks=true,bookmarks=true,%
     pdfborder={0 0 0},linkcolor=blue,urlcolor=red,%
     breaklinks=true,bookmarksnumbered=true]{hyperref}
\usepackage{wasysym}
\usepackage{times}

\def \x{\bm{x}}

\def \S{\bm{S}}
\def \s{\bm{s}}

\def \k{\bm{k}}

\def \pt{\partial}

\def \sgn{\text{sgn}}

\def \beq{\begin{equation}}
\def \eeq{\end{equation}}
\def \beqarr{\begin{eqnarray}}
\def \eeqarr{\end{eqnarray}}
\def \bspt{\begin{split}}
\def \espt{\end{split}}
\def \bef{\begin{figure}}
\def \enf{\end{figure}}

\def \bpm{\begin{pmatrix}}
\def \epm{\end{pmatrix}}

\def \si{\sigma}
\def \sib{\bar{\sigma}}

\def \Torder{\mathcal{T}}

\def \imm{\Im m}
\def \tJ{$t$-$J$ }

\long\def\symbolfootnote[#1]#2{\begingroup               %use symbols to
\def\thefootnote{\fnsymbol{footnote}}\footnote[#1]{#2}   %mark the footnote
\endgroup}                                               %rather than numbers
\newcommand{\ev}[1]{\mbox{$\langle #1 \rangle$}}
\newcommand{\dev}[1]{\mbox{$\langle\langle #1 \rangle\rangle$}}

\newcommand{\ket}[1]{\mbox{$| #1 \rangle$}}

\newcommand{\ua}{\uparrow}
\newcommand{\da}{\downarrow}

\newcommand{\eqdisp}[1]{Eq.~(\ref{#1})}
\newcommand{\refdisp}[1]{Ref.~[\onlinecite{#1}]}
\newcommand{\figdisp}[1]{Fig.~(\ref{#1})}

\def \dwx{\color{black}}
\newcommand{\ry}[1]{\textcolor{black}{#1}}

\begin{document}

%\preprint{APS/123-QED}

\title{Dynamical $t/U$ Expansion of \ry{the} %a
Doped Hubbard Model
}% Force line breaks with \\
%\thanks{A footnote to the article title}%

\author{Wenxin Ding$^{1,2}$}\email{wenxinding@gmail.com}
\author{Rong Yu$^3$}
 \affiliation{%
$^1$School of Physics and Material Science, Anhui University, Anhui Province, Hefei, 230601, China\\
$^2$   Kavli Institute for Theoretical Sciences,
University of Chinese Academy of Sciences, Beijing, China\\
$^3$Physics Department and Beijing Key Laboratory of
Opto-electronic Functional Materials and Micro-nano Devices, Renmin University, Beijing 100872, China
 }%

%\collaboration{MUSO Collaboration}%\noaffiliation
\date{\today}% It is always \today, today,
             %  but any date may be explicitly specified

\begin{abstract}
  %In this work, we first
  \ry{We construct} a new $U(1)$ slave spin representation for the single band Hubbard model in the large-$U$ limit. %. This new representation
  \ry{The mean-field theory in this representation} is more amenable to describe \ry{both the spin-charge-separation physics of the %separated
  Mott insulator} %insulating phase
  at half-filling %yet still capable of describing
  \ry{and the strange metal behavior at finite doping.} %the doped phase at the mean field level as the previous construction in which the hopping term is turned into a transverse field.
By employing a %recently developed
dynamical Green's function theory \ry{for %quantum
slave spins,
we %treat transverse field at small finite dopings perturbatively to
%compute the slave spin dynamical Green's functions. This further yields the local
calculate the single-particle spectral function %for
of electrons, and the result} %which
is comparable to %those computed by
\ry{that in} dynamical mean field theories.
 %Then, we
 \ry{We then} formulate a dynamical $t/U$ expansion for the \ry{doped} Hubbard model \ry{that reproduces the mean-field results at the lowest order of expansion. To the next order of expansion, it naturally yields an effective low-energy theory of  %which reduces the Hubbard model to
 a $t-J$ %-type low energy effective theory for the
 model for spinons self-consistently coupled to %and
 an $XXZ$ %-type effective theory
 model for the slave spins.}
% This expansion produces the mean field results at the lowest order, but enable us to further compute the spin-spin interactions.
%\textcolor{red}{SHALL WE REMOVE THE FOLLOWING SENTENCE SINCE IT READS REDUNDANT: Despite the simplicity of the mean field theory, we find that, for the single particle spectral functions, both the pole structures and the corresponding spectral weight distributions are in quantitative agreement with established lore of the field.}
%For the spin-spin interactions, we
\ry{We %find
show} that the superexchange $J$ is renormalized by doping, in agreement with the Gutzwiller approximation.
% as the well-accepted phenomenlogical proposal in Phys. Rev. B, {\bf 37}, 3759 (1988).
%In addition,
\ry{Surprisingly, we} find a new {\it ferromagnetic} channel \ry{of} %for
exchange interactions which %is {\it ferromagnetic} and not $\propto U^{-1}$.
\ry{survives in the infinite $U$ limit, %which is . %and connects to the Nagaoka ferromagnetism.
% We interpret this new term as the
as a manifestation %for
of the Nagaoka ferromagnetism.} %at finite $U$.
%  Based on these calculations, { we obtain effective dynamical theories at $\mathcal{O}(t^2/U)$ for both the spinons and the slave spins which are linked by self-consistency conditions.}
  %More interestingly, the dynamical Green's function of the slave spins mean field solution indicates possible pairing-type interactions solely stems from quantum fluctuations of the doping driven Mott-insulator-to-metal transition.
\end{abstract}

%\pacs{Valid PACS appear here}% PACS, the Physics and Astronomy
                             % Classification Scheme.
%\keywords{Suggested keywords}%Use showkeys class option if keyword
                              %display desired
\maketitle
%\tableofcontents

%%%%%%%%%%%%%%%%%%%%%%%%%%%%%%%%%%%%%%%%%%
%\section{Introduction}
{\it Introduction.~~}
\ry{Although the parent cuprates are genuine three-band, charge transfer insulators\cite{Zaanen1985}, it is widely accepted that the high-$T_c$ superconductivity\cite{Bednorz1986b} and associated anomalies in the normal state
can be adequately described within a doped single-band Hubbard model (HM)\cite{Hubbard1963} with a large onsite repulsion $U$ %~\cite{Lee2006} %A further simplified model
and %is
its descendent low-energy effective model, the \tJ model\cite{Zhang1988a,Dagotto1994,Lee2006}.}

%The cuprate high-$T_c$ superconductors\cite{Bednorz1986b} are generally considered as doped Mott insulators\cite{Lee2006} despite that the parent compounds are genuinely three-band, charge transfer insulators\cite{Zaanen1985}. It is widely accepted that a single band Hubbard model (HM)\cite{Hubbard1963} with a large onsite repulsion $U$ is an adequate description for most common experimental phenomena. A further simplified model is its descendent low-energy effective model, the \tJ model\cite{Zhang1988a}.%, which is expressed in terms of Gutzwiller-projected fermions with hopping amplitude $t$ and antiferromagnetic (AFM) superexchange interactions ($J$).

%Although the %connection between
\ry{The \tJ model %can be
is usually obtained from the HM by performing a Schrieffer-Wolff-type transformation\cite{Schrieffer1966,MacDonald1988}, which} %and HM can be established through the application of a Schrieffer-Wolff-type transformation\cite{Schrieffer1966,MacDonald1988}, which
can be considered as \ry{a version of} $t/U$ expansion. \ry{However,} it is known that certain aspects of the HM %is
\ry{are} not captured by the \tJ model. For example, the Nagaoka ferromagnetism\cite{Nagaoka1966} \ry{and %the possibility of
its possible survival %for
at %sufficiently
large-but-finite-$U$ %in thermodynamic
limit} is still under debate. Moreover, \ry{to be consistent with experimental observations,} both the hopping amplitude and superexchange interaction %strength
\ry{in the \tJ model} have to be rescaled \ry{by the doping level $\delta$} empirically, %as a function of doping level $\delta$ in order to describe experimental observations which is
known as the Gutzwiller approximation\cite{Zhang1988}. %It is also found that
\ry{Even though,} certain dynamical effects %cannot be
\ry{are not} captured %in
\ry{within} the standard Gutzwiller-projected \tJ model\cite{Delannoy2005, Anderson2006, Lee2006, Phillips2010}
%(After the transformation, the new fermion operators are not {\it local} in terms of the original fermions. ?).

%Even for the simplified \tJ model, the implementation of Gutzwiller-projection prove to be difficult. typically can only be done on a trial wavefunction numerically\cite{}, or in the infinite-dimensional limit where the dynamical mean field theory (DMFT) becomes exact\cite{}.

%The theory capable of quantitatively describing the low energy dynamical effects of Gutzwiller-projection in the \tJ model, known as the {\it extremely correlated Fermi liquid} (ECFL) theory, is only achieved recently by Shastry\cite{Shastry2011,Shastry2010,Shastry2013} via the Schwinger's equations-of-motion (SEoM) approach\cite{Schwinger1953,Schwinger1954,Schwinger1954a,Fields1954}.

%\refdisp{Pairault2000} formulated a (self-consistent) $t/U$ perturbation theory at finite temperatures. The results works for half-filled case, and probably for doped case at high temperatures. At high temperatures, the spectral weights behave as those of a band insulator. The theory breaks down for low temperatures. * spectral functions of raw perturbation theory  also shows negative spectral functions/LDOS.

%\refdisp{Adibi2018} solves for disordered Mott insulator at half-filling.

%Theoretically, a more ideal approach for extracting the
\ry{A more formal approach to construct a low-energy effective theory %of
for the} HM would be from either a dynamical $t/U$ perturbative series \ry{expansion} in terms of Green's functions\cite{Furman1988,Pairault2000} or %effective theory obtained from
a cumulant expansion\cite{Ding2014}. However, %this
\ry{either} is generically difficult due to the {\it noncanonical} nature of the atomic basis when doping is finite. % which one needs to use as the unperturbed theory.
%Although, as shown by B. S. Shastry, it
\ry{It is} possible to describe the dynamics of noncanonical theories via the Schwinger's-equations-of-motion (SEoM) \ry{approach~\cite{Shastry2011}, but} it is %not known
\ry{still unclear} how %such
\ry{to do the} $t/U$ expansion %can be done for
\ry{using} the exact SEoM %of
\ry{of the} HM. A %simpler
\ry{feasible} way would be to start with \ry{a %``good''
slave-boson-type representation
that can both faithfully reproduce the local single-particle spectrum of the HM and fulfill the spectral sum rule of Green's functions.}
% , which is deeply connected to the algebraic structures of the Hubbard operators, and
\ry{Then} use the path integral method to dynamically integrate out the high energy degrees of freedom %and
to obtain the corresponding low energy effective theory. %By ``good'', we mean the representation should not only faithfully reproduce the local single-particle spectrum of the HM, %on one site,
%but its Green's functions \ry{should} also satisfy a spectral sum rule.
%Previously such
\ry{Such} a dynamical theory has been obtained at half-filling via a slave rotor representation~\cite{Ding2014}. \ry{Unfortunately it cannot be applied to finite doping due to the limited Hilbert space} {\dwx of the rotors}. %which only works at half-filling\cite{Ding2014}.
On the other hand, the $U(1)$ slave spin representation\cite{Yu2012e} \ry{has a %holds the potential to fulfill the requirement for dynamical $t/U$ expansion in the doped regime due to its
larger Hilbert space and would be suitable for the dynamical $t/U$ expansion %in the doped
at finite doping. But a major obstacle lies in the redundant spin $SU(2)$ symmetry in both the slave spin and fermionic spinon sectors in its conventional construction.} %However, the $SU(2)$ spin operators themselves are also noncanonical, and poses as a major obstacle.

In this work, we %construct
\ry{introduce} a new $U(1)$ slave spin representation %which is more suitable for describing
\ry{that better describes} the spin-charge \ry{separation in a Mott insulator. %separated Mott-insulating
In light of this convenient representation, we construct %construction of the
a dynamical $t/U$ expansion for the doped HM by employing a perturbative SEoM theory for the %$SU(2)$
slave spins.}
%limit comparing to the previous constructions, and then employ a recently developed, perturbative SEoM theory for the $SU(2)$ spin-$1/2$ operators to explore the dynamical $t/U$ expansion for the doped HM in regimes where $\delta \ll 1$. The rest of this work is organized as the following.
%We first introduce the new $U(1)$ slave spin representation. Then we introduce the SEoM approach for $SU(2)$ spin operators and apply it to the atomic limit of the HM and compute the superexchange interaction strength. Next, we move to the mean field theory at finite doping. We treat the hopping term of the slave spin sector with a Weiss mean field approximation, and calculate the dynamical slave spin Green's functions perturbatively.
This enables us to calculate electronic spectral functions, which agree with CDMFT\cite{Sakai2009} results and ECFL\cite{Shastry2011} results obtained by SEoM for \tJ model ($U \rightarrow \infty$)% and other previous results\cite{}
on the pole structure and spectral weight distribution, despite the mean field nature of %the
\ry{our} theory. We \ry{further} %also
compute the %superexchange
\ry{spin-spin interaction strength %subsequently, and propose an effective theory at the first
up to the $O(t/U)$ order of %$t/U$
the expansion beyond %on top of
the saddle point. An effective \tJ model with rescaling factors agreed with those in Gutzwiller approximation naturally emerges in our theory. In addition to the antiferromagnetic superexchange coupling we find a ferromagnetic exchange coupling surviving up to $U\rightarrow\infty$, which connects to the Nagaoka ferromagnetism. %theory. We find that the phenomenological rescaling factors for \tJ model naturally emerges out of our dynamical calculations. In the end, we
We finally} discuss the implication of %this work
\ry{our theory and %the
future} prospect of our new approach to the HM.

%%%%%%%%%%%%%%%%%%%%%%%%%%%%%%%%%%%%%%%%%%
% \section
{\it $U(1)$ slave spin representation of the Hubbard model.~~}
We %introduce a $U(1)$ slave-spin representation of
\ry{rewrite} the physical electron operators $d_{i\sigma}$ and $d_{i\sigma}^\dagger$ as
    \begin{align}
      d_{i\sigma}^\dagger =(S^+_{i a} + S^+_{ib}) f^\dagger_{i \sigma}/\sqrt{2}, ~
      d_{i\sigma} =  (S^-_{ia} + S^-_{ib}) f_{i\sigma}/\sqrt{2}, \label{eq:slave-spin-new-rep}
    \end{align}
   % {\dwx Normalization seems wrong in the free-limit? Instead, put $\frac{1}{\sqrt{1-(S^z_{i,tot})^2}}$? in order guarantee physical fermion spectral weight sum rule?}
   \ry{where $S^{\pm}_{i a}$ are ladder operators of $S=1/2$ slave spins, and $f_{i \sigma}$ is a fermionic spinon operator.}
    In contrast to previous
    constructions~\cite{Yu2012,deMedici2005}, %where $a$ and $b$ refer to two spin flavors,
    in this representation, the slave-spin %index is
    \ry{indices $a$ and $b$ are} no longer associated with the physical spin index $\sigma$, \ry{so that the slave spins and spinons respectively carry the charge and spin degrees of freedom,} indicating a full charge-spin separation. The constraint becomes
      $\sum_{l=a,b} S^z_{i s}  = \sum_{\si = \ua, \da} (f^\dagger_{i \si} f_{i \si} - \frac{1}{2}),$
    in contrast to previous constructions, in which the constraint is for each spin flavor.
    The \ry{Hamiltonian of the %Hubbard-$U$ interaction remains unchanged, so that the full
    single-band %Hubbard model
    HM in this slave-spin representation} is written as
    \begin{align}
      &  H = H_{S,0} + H_{f,0} + H_{t}, \label{eq:H-slave-spin}
    \end{align}
    with
   $  H_{S,0} = \frac{U}{2} \sum_{i} (\sum_{s} S_{i s}^z)^2 + h \sum_{i s} S^z_{i s},~
        H_{t} =- \sum_{ij \si}  \frac{t_{ij}}{2} (S^+_{i a} + S^+_{i b})(S^-_{j a} + S^-_{j b}) f^\dagger_{i \si} f_{j \si},~
       H_{f,0} = (- h - \mu) \sum_{i \si} n^f_{i\si}.$
  Here $\mu$ is the chemical potential and $h$ is a Lagrangian multiplier %for implementing
  \ry{to implement} the constraint.

 As we shall demonstrate later, this representation reproduces the Green's function of a Mott insulator %via
 \ry{in} a slave rotor representation\cite{Florens2002,Florens2004} which has been shown to have captured the impurity physics %in
 of a Mott insulator\cite{Ding2018}. %This new slave-spin representation has
 \ry{This representation hence provides} a better description of the %spin-charge separation physics for the half-filled
 Mott insulating phase \ry{at half-filling. %In the meantime , this representation
 Meanwhile, at the mean-field level it} retains the capability to describe \ry{the %doped
 phases at finite doping} %at mean field level
 just as in previous works\cite{Yu2012e}.
 %However,
 \ry{In this work, we %shall
 focus on %obtaining
 the dynamical properties %of the doped phase described by the slave-spin theory
 and %the higher order
 the low-energy effective theory obtained beyond the mean-field level at finite doping %in which those
 where dynamical fluctuations are taken into account.}

 %%%%%%%%%%%%%%%%%%%%%%%%%%
 {\it The atomic limit.}~ To %consider
 \ry{study} the dynamical properties of %the
 slave spins, we use the \ry{SEoM %Heisenberg-Schwinger equations of motion
 method, which converts the} %exact
 Heisenberg equations of motion of the operators into exact equations of motion of the Green's functions or propagators of these operators. \ry{This is formally done via a perturbation theory on an effective XXZ %-type
 spin model under transverse and longitudinal fields which is similar to those developed for Heisenberg models~\cite{Kondo1972,Shimahara1991a,Gasser2001,FROBRICH2006,Nolting2008,Majlis2014}. The details on the perturbation approach will be presented elsewhere~\cite{Ding2019b}, and here for simplicity, we only show main results.} %Similar approach of Heisenberg spin Green's functions has been studied in \cite{Kondo1972,Shimahara1991a,Gasser2001,FROBRICH2006,Nolting2008,Majlis2014}, but without emphasizing on an exactly solvable limit. A perturbative theory via the Heisenberg-Schwinger equations of motions, which is the utilized throughout this work, is developed by one of the authors and presented independently elsewhere.

The slave spin Green's function %$G^{\alpha \bar{\alpha'}}_{\eta, S s s'}[i,f]$
is defined as
\begin{align}
G^{\alpha \bar{\alpha'}}_{\eta, S, s s'}[i,f] =\ev{\mathcal{T}[S_{i s}^\alpha (t_i) S_{f s'}^{\bar{\alpha}'} (t_f)]_{\eta}} - c_\eta \ev{S_{i s}^\alpha} \ev{S^{\bar{\alpha}'}_{f s'}},
\end{align}
with $\alpha = +$ or $-$, $s = a$ or $b$, $\eta = B \text{ or } F$ representing the sign of the time-ordering and $c_B = 1$ and $c_F = 0$.% which is to subtracts the static components.
Throughout this work, we use the labels $[i,f]$ to denote \ry{the space-time coordinates} %and time coordinates pairs
of \ry{the} initial and final states $[\x_i, t_i; \x_f, t_f]$.
To simplify the notation, from now on we %shall
drop the slave spin index $s$ so that $G^{\alpha \bar{\alpha'}}_{\eta,S}$ %and $G^{\alpha \bar{\alpha'}}_{S\pm}$
indicates a form of $2\times 2$ matrix in spin space. \ry{One may freely choose to use either bosonic ($B$) or fermionic ($F$) Green's function in the calculation because each will give a complete set of equations. %which shall be represented in terms of Pauli matrices $\si_i$s. Although we need to employ both $B$ and $F$ Green's functions in the calculation for different purposes, the electronic properties only explicitly involves $G^{\alpha \bar{\alpha'}}_{B, S s s'}$. Therefore,
Here we %will
only} show results for $G^{\alpha \bar{\alpha'}}_{B, S s s'}$ \ry{and present the form of %Expressions for
$G^{\alpha \bar{\alpha'}}_{F, S s s'}$ %will be shown
in SM} as needed.

In the \ry{atomic limit (corresponding to Ising slave spins)}, we can obtain the exact dynamical Green's function~\cite{SM,Ding2019b}. An arbitrary state (not necessarily \ry{an} eigenstate) of $H_{S,0}$
can be fully characterized by the set of parameters $(M,~ m, ~\Delta m^2)$, \ry{where} %which are defined as
%\begin{align}
%  \begin{split}
 $ M=\ev{S^z_{a} + S^z_{b}},\
  m=\ev{S^z_{a} - S^z_{b}},\\
  \Delta m^2 = \ev{(S^z_{a} - S^z_{b})^2}.$
%\end{split}
%\end{align}
% Zero-T limit solution is still determined by minimize ground state energy in order to determine $\ev{S^z_{s}}$.
\ry{Solving $H_{S,0}$ at half-filling (see Supplemental Material (SM) and \refdisp{Ding2019b})} we find
$
  h=0, \quad M = 0, \quad m^2 + \Delta m^2 = 1.
$
\ry{Here the} uncertainty of $m$ and $\Delta m^2$ %is related to
\ry{reflects} the spin degeneracy %of
\ry{at} the atomic limit. Choosing $m = 1$, we \ry{get}~\cite{SM,Ding2019b} %find
\begin{align}
  G^{\alpha \bar{\alpha'}}_{B,S} = \frac{ \alpha  \delta_{\alpha \alpha'} \si_z}{\omega - \alpha \si_z U/2}.\label{eq:atomic-half-filling-GS}
  % G_{S\sib} = \frac{-1}{\omega - U/2}.
\end{align}
  Switching to imaginary frequency $\omega \rightarrow i \nu$, we can recover the slave-rotor %rotor's
  Green's function $G_{X}[\nu]$ at half-filling $ (G^{+-}_{B, S, a a} + G^{+-}_{B, S, b b})/2 \rightarrow G_{X}[\nu] =(\nu^2/U + U/4)^{-1}$, %which is
  consistent with previous works\cite{Florens2002,Florens2004,Ding2014}.

% As we show in Supplement Materials(SM) and \refdisp{Ding2019b},

   %%%%%%%%%%%%%%%%%%%%%%%%%%%%%%

 {\it %Saddle point
 Effective theory at saddle-point level}.~~
   Following the construction of \refdisp{Ding2014}, \ry{when the hopping is turned on,} at the saddle point level the theory is decoupled into an effective slave spin theory
   \begin{align}
     & H_{S,eff} = H_{S,0} +  H_{S, t}, \label{eq:H-S-eff-0-order}
   \end{align}
  with $ H_{S,t} = - \sum_{ij, s s'} (Q_{f,ij} S^+_{is} S^-_{js'} + h.c.)$, and an effective $f-$spinon theory
   \begin{align}
     H_{f,eff} =  H_{f,0} + H_{f,t} \label{eq:H-f-eff-0-order}
   \end{align}
   with  $H_{f, t} = - \sum_{ij \si} (Q_{S,ij} f^\dagger_{i \si} f_{j \si} + h.c.)$.
    %which
    \ry{The parameters}
    %\begin{align}
      $Q_{f,ij} = \sum_{\si} t_{ij} \ev{f^\dagger_{i\si} f_{j\si}}, \quad
     Q_{S,ij} = \sum_{s s'} t_{ij}\ev{S^+_{is} S^-_{js'}}$ \ry{are %linked/coupled through the
     self-consistently determined.} %parameters
   %\end{align}
%%%%%%%%%%%%%%%%%%%%%%%%%%%%%%%%%%%%%%%%%%%%%%%%%%%%%%%%%%%%%%%%%%%%%%%%%%%%%%%%%%%%%%%
% STOP! STOP! STOP!
%%%%%%%%%%%%%%%%%%%%%%%%%%%%%%%%%%%%%%%%%%%%%%%%%%%%%%%%%%%%%%%%%%%%%%%%%%%%%%%%%%%%%%%
\ry{In this theory the quasiparticle spectral weight is defined as $Z={\ev{S_x}}^2$. The Mott insulator at half-filling is described by a paramagnetic state of the slave spins with $\ev{S_x}=0$. Doping the Mott insulator drives the system to a metallic state, in which the slave spins form long-range order $\ev{S_x}\neq 0$.  %At finite doping $\delta$ and
When only the nearest neighbor (nn) hopping is taken into account, $ Q_{f,ij}$
is determined by %but only on
the spinon density %which
(that equals the electron density) and %does not
is independent of $Q_{S,ij}$. With this, the self-consistency at the saddle point
%theory
is trivially achieved at finite doping $\delta$.} %and we find $Z\propto\delta\simeq Q_{S,ij}$. %When the system becomes metallic, i.e. the slave spin develops long-range order which can be chosen to be along $S_x$,  $Q_{S,ij} \simeq \ev{S_x}^2 = Z \propto \delta$.
 %considered taken into account,  Therefore, self-consistency of the saddle point theory is trivially achieved even at finite doping.
For simplicity, in the rest of this work, we %shall
restrict to the cases %of
$\delta \sim 0$ where a perturbation theory is presumably valid. %, instead of the full domain of mean field theory obtained by diagonalization.

   %%%%%%%%%%%%%%%%%%%%%
%{\it Weiss mean field theory of the hopping term}.~~
\ry{At the saddle-point level, the slave-spin Hamiltonian can be solved by implementing a Weiss mean-field approximation to $H_{S,t}$, which has been widely adopted in previous works for single- and multi-orbital systems\cite{DeMedici2005a,Hassan2010,Yu2010,Yu2012e, leewc}. In this approximation}

%   In previous works\cite{DeMedici2005a,Hassan2010,Yu2010,Yu2012e}, the slave spin sector of saddle point theory was solved at a Weiss mean field level, which already contributes significantly towards the understanding of iron-based superconductors.
%   In this section, we solve $H_{S,eff}$ for finite doping at mean field level and compare with Weiss mean field results obtained by exact diagonalization. The Weiss mean field approximation is to decouple $H_{S,t} \rightarrow H_{S,tMF} $ as
   \begin{align}
     %\begin{split}
       %& H_{S,t} \rightarrow \\
       H_{S,t}\approx H_{S,tMF} = -  \sum_{is} \big( (\sum_{\ev{ij} s'} Q_{f,ij} \ev{S^-_{js'}}) S^+_{is} + h.c. \big).
   %\end{split}
   \end{align}
\ry{The $U(1)$ symmetry of the slave spins is broken, and we choose }
%   Since the emergence of $\ev{S_{i s}^{\pm}} \neq 0$ is from spontaneous-symmetry-breaking, we can choose the direction of the magnetization at our convenience:
$\ev{S^+_{is}} = \ev{S^-_{is}} = \ev{S^x_{is}} = M_x$. On a 2D square lattice with only  nn hopping, $Q_{f,ij} = Q_f$ \ry{so that the mean-field Hamiltonian $H_{S,MF} = H_{S,0} + H_{S,tMF}$ becomes}
%\begin{align}
  %\begin{split}
%    & H_{S,MF} = H_{S,0} + H_{S,tMF}, %\\
    %& = \sum_{i,s} \Big( \frac{U}{2} S^z_{is} S^z_{i\bar{s}} + h S^z_{is}  - D M_x Q_f S^x_{is} \Big),
%\end{split}
%\end{align}
%where we consider only nn hopping and assume uniform values for $Q_{f,ij} = Q_f$, $D=4$ is the number of nn bonds, and $M_x$ is to be determined self-consistently from solving $H_{S,MF}$. This is
a local Hamiltonian. %and
\ry{It} can be solved by exact diagonalization \ry{and one finds $Z\propto\delta\simeq Q_{S,ij}$. The same result can be alternatively arrived from the lowest-order Green's function $G^{\alpha \bar{\alpha'}}_{B,0}[\omega]$ in our dynamical perturbation theory  %We compare calculations of our Green's function perturbation theory
in the limit of small doping $\delta$ (see SM). }%The mean field results is shown and discussed in detail in SM. Here we focus on the dynamical aspect of the theory.

In the presence of perturbations, the lowest order effect is that the ground state $\ket{\psi(M,~ m, ~\Delta m)}$ is  renormalized. Hence, %we first consider the exact Green's functions
$G^{\alpha \bar{\alpha'}}_{B,0}[\omega]$ for arbitrary states under the evolution of $H_{S,0}$ \ry{takes the following form}:
\begin{align}
  \begin{split}
    & G^{\alpha \bar{\alpha}}_{B,S,0}[\omega] =  \si_0 \frac{ X (\omega + \alpha h) + (M + \alpha (1-X)) U/2}{(\omega + \alpha h)^2 - U^2/4}, \label{eq:G_B0}
\end{split}
%\\
%  \begin{split}
%   & G^{\alpha \bar{\alpha}}_{F,S,0}[\omega] = \frac{ (\omega + \alpha h) \si_0 + \alpha (M \si_0 ) U /2}{(\omega + \alpha h)^2 - U^2/4},
%\end{split}
%\label{eq:G_F0}
\end{align}
where $X = 1 - \Delta m^2$. In the above expression, we already %make
\ry{made} use of the following properties: i) $m = 0$ in the onset of transverse field; ii) $\Delta m^2 \simeq 1$ near the doping-driven-Mott-insulator-to-metal-transition (dMIMT), so that $X$ is also a small parameter. Numerical calculations find $X = c_0 \delta$ in this mean field approximation where $c_0$ runs from about $1.3$ near $U_c$ to $1$ for $U\rightarrow \infty$.

%The transverse magnetization, i.e. the quasiparticle weight, the magnetization, i.e. the hole density, and $\Delta m^2$ correction, to the lowest order in $h_x$, are found to be
%\begin{align}
%  \ev{S^x_{a}} \simeq h_x \left[\frac{U \si_0 - 2 h M \si_0 }{U^2 - 4 h^2}\right]_{aa},\\
%\delta =  M = 2 \ev{S^z_a} \simeq \frac{- 4 h h_x^2}{U (U^2/4 - h^2)},\\
%  \delta \Delta m^2 \simeq \frac{- h_x^2}{U^2/4 - h^2}.
%\end{align}
%which leads to $Z  = M_x^2 \propto \delta$ by solving the self-consistency equation
%\begin{align}
%h_x = D M_x Q_f.
%\end{align}
%
%The complete self-consistent solution of the Weiss mean field theory for the hopping terms is done via exact diagonalization. The mean field parameter results are shown in Supplement Materials. We obtain a second order Mott-insulator-to-metal transition at finite $-\mu_c = h_c \simeq 2.8? $ for $U=10$. $Z = M_x^2 = \delta$.
%

{\it Perturbative correction to slave spin Green's functions.~~}
Consider the perturbation to the slave spin Green's functions of $H_{S,MF}$ as the following:
\begin{align}
  G^{\alpha \bar{\alpha'}}_{B,S}[\omega] \simeq G^{\alpha \bar{\alpha'}}_{B,S,0}[\omega] + G^{\alpha \bar{\alpha'}}_{B,S,1}[\omega], \label{eq:G-BS-1st-order}
\end{align}
where $G^{\alpha \bar{\alpha'}}_{B,1}[\omega]$ is the lowest order correction of $G^{\alpha \bar{\alpha'}}_{B}[\omega]$ (other than the change in wavefunction under the evolution of $H_0$).
The SEoM theory gives
\begin{align}
  \begin{split}
    & G^{\alpha \bar{\alpha'}}_{B,S,1, s s'}[\omega] = \frac{\alpha}{\omega + \alpha h} \Big( \frac{\alpha' (U \ev{S^x_{s}} + h_x) I_x}{2}  \\
    &  + \sum_{\alpha''} \frac{\alpha'' h_x}{2 \omega} ( U \ev{S^x_{\si}}   G^{\alpha'' \bar{\alpha'}}_{B,S,0, \bar{s} s'}[\omega]  + h_x G^{\alpha'' \bar{\alpha'}}_{B,S,0, s s'}[\omega])\Big).
  \end{split}
\end{align}
Although the bare first order perturbation to $G^{\alpha \bar{\alpha'}}_{B,S}$yields correct results for observables, like second order correction to $\ev{S_z}$ etc., it violates the spectral weight sum rule significantly, especially when $U$ is large. \ry{In principle, this %This is
can %presumably %can
be fixed by going} %when we go
to higher order perturbations. %theory which is still under development.
%To remedy this issue within the scope of this work,
\ry{And here we adopt a simple} %make a
random-phase-approximation (RPA) to the diagonal (in both the slave spin flavor index and the $\alpha$ superscript index for which $+-$ or $-+$ is considered diagonal) components of $G^{\alpha \bar{\alpha'}}_{B,S}$:
\begin{align}
  G^{+-(-+)}_{B,S,ss}[\omega] \simeq \Big(G^{+-(-+)}_{B,S,0,ss}[\omega]^{-1} - G^{+-(-+)}_{B,S,1,ss}[\omega]\Big)^{-1},
\end{align}
whereas the off-diagonal components are left the same. \ry{As we will show in below,
%Although
the spectral sum rule approximately holds by taking this RPA form of Green's functions at small dopings.} %is not proved via higher order perturbation theory, which is not trivial for noncanonical theories, it is still a legitimate small parameter ($M_x$) (inverse) expansion for the diagonal terms since $G^{+-(-+)}_{B,S,0,ss} \neq 0$.

{\it The electronic spectral functions}.~~
\ry{Knowing the Green's function of slave spins, the single-electron spectral function at finite doping is readily calculated. Here we discuss the local spectral function of electrons within the above mean-field approach.
%In this section, we consider the electronic spectral functions of the mean field theory for finite dopings.
%According to
From \eqdisp{eq:slave-spin-new-rep}, the electronic %spectral functions can be approximately expressed as
Green's function} $i G_d[i,f] \simeq (\sum_{s s'} i G^{-+}_{B, S s s'}[i,f]) (i G_{f}[i,f])$, which %can be transformed to
leads to the electronic spectral function
%\ev{\mathcal{T} \big[ d_{i}[t_i] d^\dagger_{f}[t_f] \big]_-} \\&
    \begin{align}
      &  \rho_d[\omega,\x] = \int d\omega' (\sum_{s s'} \rho^{-+}_{B, S s s'}[\omega + \omega', \x]) \rho_f[\omega'], \label{eq:electron-spectral-func}
\end{align}
where the slave spin spectral function $\rho^{-+}_{B, S s s'}[\omega, \x] = - \pi^{-1}\imm [G^{-+}_{B, S s s'}[\omega - i \eta^+ \sgn[\omega]]]$.
%In
\ry{At the mean field %case
level, we %can consider
take} $\rho_f[\omega'] \simeq (\pi)^{-1} \delta(\omega')$, %. Therefore,
\ry{and} \eqdisp{eq:electron-spectral-func} becomes
\begin{align}
  \rho^{MF}_d[\omega, \x] \simeq \sum_{s s'} \rho^{-+}_{B, S s s'}[\omega, \x].
\end{align}
%For such mean field theory,
Note that the slave-spin Green's function in \eqdisp{eq:G-BS-1st-order} is also local, %only has local components, i.e.
so that $\rho^{MF}_d[\omega, \x] %\rightarrow
= \rho^{MF}_d[\omega]$. %From here on, we drop the spatial coordinates and all functions/observables considered below are local quantities.

\begin{figure}
  \centering
  \subfigure[]{\label{subfig:rho-d}\includegraphics[width=0.48 \columnwidth]{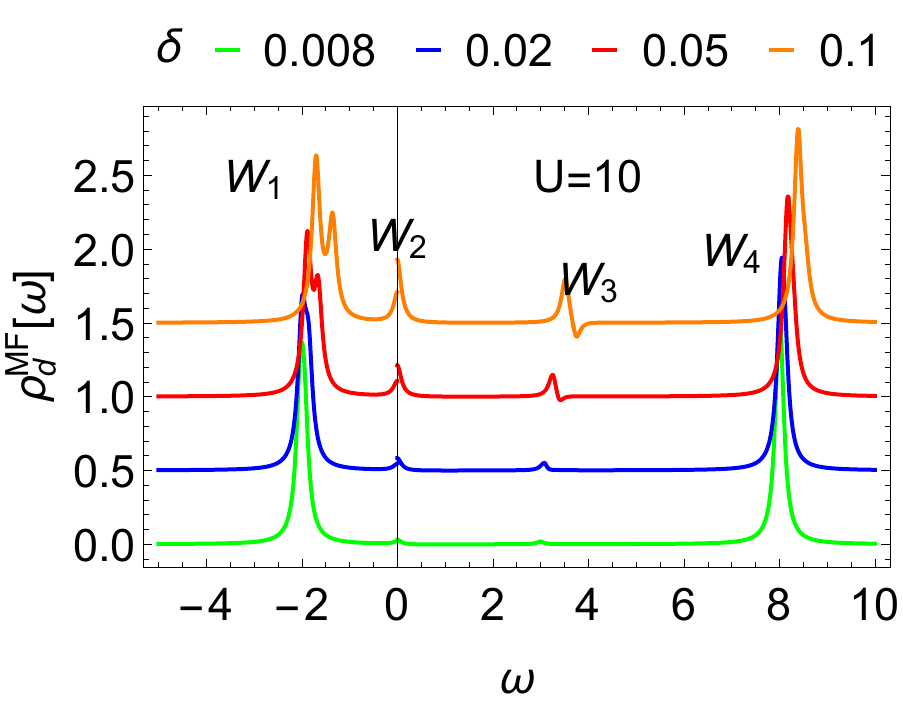}}
  \subfigure[]{\label{subfig:wplot}\includegraphics[width=0.48 \columnwidth]{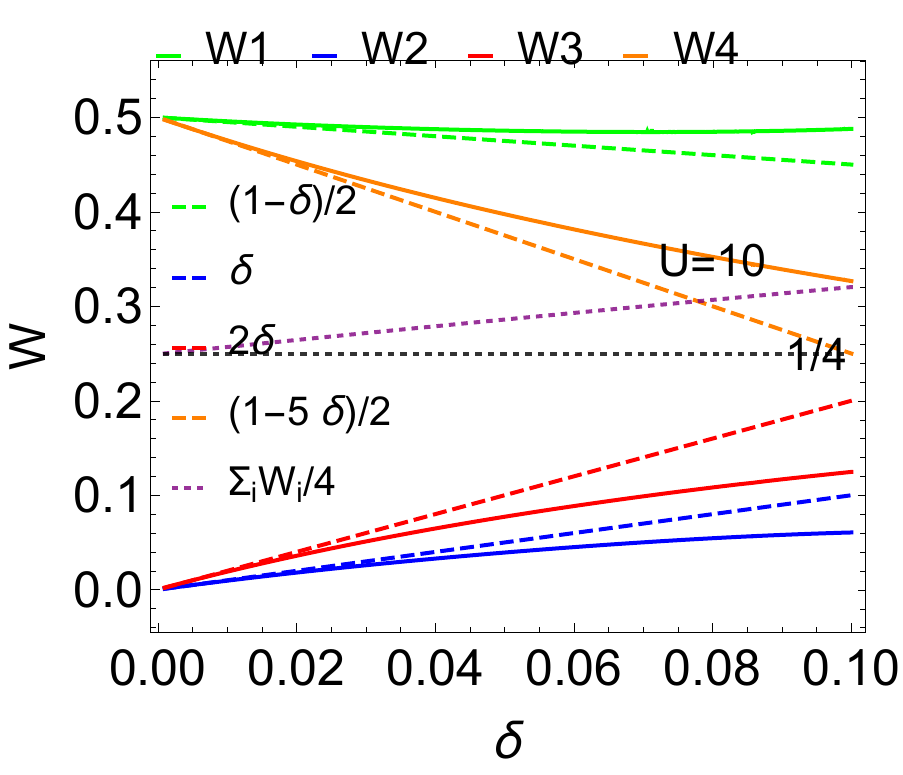}}
  \caption{(Color online) (\ref{subfig:rho-d}) %the
  local electronic spectral functions %plotted
  for $\delta = 0.008,~0.02,~0.05,~0.1$ with a $0.5$ shifting for each curve; (\ref{subfig:wplot}) the spectral weights for each pole %plotted
  as a function of $\delta$ and compared with the corresponding asymptotic behaviors. %The
  Together shown is the average spectral weight $\sum_i W_i/4$ of the poles, which indicates approximate conservation of the total spectral weight (see text).
  %total spectral weight $\sum_i W_i/4$ (purple dotted line) is preserved up to an error $\ll \delta$, which can be told by comparing to the constant $1/4$ (black dotted line). %The
  A Lorentzian broadening factor %is set to
  $\eta^+ = 0.1$ is used in each curve %for all $\delta$s
  in  (\ref{subfig:rho-d}),
  and $\eta^+ = 0.01$ is used %. But
  in (\ref{subfig:wplot}). %, we set $\eta^+ = 0.01$ for computing the spectral weights $W_i$s.
  }
  \label{fig:rho-d}
\end{figure}

We show the evolution of the \ry{local} electron spectral function with doping level $\delta$ in \figdisp{subfig:rho-d}, from which we can identify %(approximately)
four distinct poles, %which are %. We
labeled %the corresponding spectral weights of these poles
as $W_i,~(i=1,\dots,4)$, respectively.
%as illustrated in \figdisp{subfig:rho-d}, and plot
\ry{The doping dependence of the calculated spectral weight at each pole $W_i$ is shown in \figdisp{subfig:wplot}.
%the computed $W_i$s as functions of doping levels along with their corresponding asymptotic behaviors in \figdisp{subfig:wplot}.
We also plot together the quarter of total spectral weight $W_{total}/4 = \sum_i W_i/4$ %in \figdisp{subfig:wplot}
which indicates the spectral weight sum rule is conserved up to an error %at the order ($n$)
of the perturbation theory $\mathcal{O}(h_x)$, much smaller than the doping level $\delta$.}

Use the bare first order perturbation results, we identify that the poles of $\rho^{MF}_d[\omega]$ are approximately located at $\omega_i \in \{h-U/2, 0, h, U/2+h\}$.
The corresponding linear fittings for the spectral weight at these poles (dashed lines) are $\{(1-\delta)/2,~\delta,~2 \delta, ~(1- 5 \delta)/2\}$. %, which are plotted as dashed lines with the same color in \figdisp{subfig:wplot}.
These results %which respectively
clearly show that the poles respectively
correspond to the $\{$%LHB
lower Hubbard band, Fermi energy, mid-gap states, %UHB
upper Hubbard band$\}$, and the spectral weight redistribution among them with doping.
The spectral weight evolution is consistent with the analysis of CDMF results in \refdisp{Sakai2009} %if we identify
\ry{when we take the doublon density %as
to be} $n_{doublon} = 2 \delta$ in our mean-field theory. %The total spectral sum is conserved up to an error at the order ($n$) of the perturbation theory $\mathcal{O}(h_x^n)$.
%\ry{SHALL WE ADD SOME DISCUSSION ON THE MID-GAP STATE $W_3$?}

An interesting and important feature \ry{in the spectrum is the asymmetric structure of the pole $W_2$ about the Fermi level $\omega=0$. Though the jump of $\rho^{MF}_d[\omega]$ across $\omega = 0$ %. This
is generally an artifact %of the approximation of
coming from treating the $f$-spinon %DOS
density of states as a $\delta$-function, the asymmetry of the spectrum is a physical %It is a
consequence of the pole} %$\omega_2 = 0$
of $G^{\alpha \bar{\alpha}'}_{B,S}$. Note  that, this is a {\it zero} frequency pole of a {\it time-ordered} Green's function. Thus, it is a dynamical pole whose imaginary part should be interpreted as $\imm [1/(\omega + i \delta \sgn[\omega])] \propto \pt_\omega\delta(\omega)$. This pole contributes a component $\propto \pt_\omega \rho_f(\omega)$  to $\rho_d(\omega)$ after the convolution. Considering that the $f$-spinons are treated as canonical fermions whose self-energy is even in $\omega$, a dynamical particle-hole odd component\cite{Shastry2012} \ry{accounting for the asymmetric spectral weight %will
should} be incorporated in $\rho_d(\omega)$ through this pole, %which is found to be
\ry{and this is} an important hallmark of strong $U$ correlations in recent microscopic theories including the ECFL theory\cite{Shastry2011,Shastry2012}, the hidden Fermi liquid theory\cite{Anderson2009,Casey2011}, as well as DMFT results\cite{Deng2013,Xu2013d}.

{\it Spin-spin interactions of the doped phase.~~}
%  {\it Superexchange interactions in the atomic limit}.~~
\ry{The dynamical expansion allows us to extract }  %The amplitudes of
the superexchange interactions in terms of the $f$-spinons
  \begin{align}
    H_{J} = \frac{J_{ij}}{2} f^\dagger_{i\alpha} f_{i\beta} f^\dagger_{j \beta} f_{j \alpha} \Rightarrow J_{ij} \S_i \cdot \S_j.
    %.J_{ij} f^\dagger_{i\alpha} f_{j\alpha} f^\dagger_{j\beta} f_{i\beta}
  \end{align}
  Here $J_{ij}$ \ry{is} %are
  obtained by contracting the vertex at one-loop level\cite{Ding2014} as
  \begin{align}
    \begin{split}
      & J_{ij} = t_{ij}^2  \int \frac{-d\omega}{2 \pi} \sum_{%\substack{
      ss' %\\
      l l'%}
      } G^{+-}_{B,S,ss'}[\omega, i,i] G^{-+}_{B,S,l l'}[\omega, j,j].
      %\\    %& = 2 t_{ij}^2 / U = J_0.
    \end{split}
        \label{eq:J-expr}
  \end{align}
  \ry{Plugging in the slave-spin Green's function at the atomic limit, %We
  we %recover the same
  find} the superexchange interaction at half-filling: $J_0 = 2 t_{ij}^2 / U$. The missing factor of 2 can be restored %similarly
  \ry{when taking into account the fluctuations %due to
  of the hopping term %is considered just
  similar as} in \refdisp{Ding2014}.
  %Using the atomic $G_{B, S s s'}$ at half-filling, i.e. \eqdisp{eq:atomic-half-filling-GS}, we recover the slave-rotor and two-site 2nd order perturbation results $J_{ij} = 4 t_{ij}^2 / U$.

%In this part, we discuss the spin-spin interactions mediated by the slave spins among the $f$-spinons.
%In the metallic phase
\ry{At finite doping, more channels of spin-spin interactions arise from different dynamical processes, and they can be calculated term by term according to} %are available, each of which is dynamically different and should correspond to different dynamical spin-spin interactions.
%Therefore, we use
\eqdisp{eq:J-expr} and (\ref{eq:G-BS-1st-order}). %to compute and distinguish different channels of spin-spin interactions as the pole structure is more transparent for \eqdisp{eq:G-BS-1st-order}. Ignoring the spatial indexes and summation over the slave spin flavor indexes, we obtain the leading spin-spin interactions as
Up to the leading orders,
\begin{align}
  J &= J^{(0)} +  J^{(1)} + \dots,
\end{align}
where $J^{(0)} = - t^2\int \frac{d\omega}{2 \pi} G^{+-}_{B,S,0} G^{-+}_{B,S,0}$, and
$J^{(1)} = - t^2\int \frac{d\omega}{2 \pi}  (G^{+-}_{B,S,1} G^{-+}_{B,S,0} + G^{+-}_{B,S,0} G^{-+}_{B,S,1}).$
  %\\J^{(2)} & = - t^2\int \int \frac{d\omega}{2 \pi} G^{+-}_{B,S,1} G^{-+}_{B,S,1}
$J^{(0)} $ is the superexchange interactions, which is still governed by the virtual transition between the LHB and UHB as $G^{+-(-+)}_{B,S,0}$ only have poles at $\omega_1$ and $\omega_4$. To the lowest order in $\delta$, we have
\begin{align}
  J^{(0)} \simeq  J^*_0 (1 - 2 \delta)
  %(\frac{1}{1 - 4 a^2} - \frac{2\delta}{1+2a}),
\end{align}
where $a = h/U$ and $J_0^*= J_0 (1-4 a^2)^{-1}$ which is the bare superexchange interaction strength if only the energy shifting of the LHB and UHB caused by doping is accounted for.
%This GA factor reflects the spectral weight shifting of the LHB and UHB.
%This factor was proposed by Zhang et. al.\cite{Zhang1988} phenomenlogically.
%However, as shown in \figdisp{subfig:J-delta}, $J^{(0)}$ increases with $\delta$. This is because the $a$-factor decreases significantly (in fact faster than $\delta$, which is shown SM) with $\delta$ which is likely an artifact of the mean field theory.
$J^{(1)}$ is due to the virtual transitions between LHB and the %new poles
mid-gap pole $W_2$. To the lowest order, we have
\begin{align}
J^{(1)}  \simeq  - \frac{M_x^2 t^2 }{1 + 4 a},
\end{align}
which is ferromagnetic. %, and does {\it not} depend on $U$.
By extrapolation, we find $M_x^2\big\vert_{U\rightarrow \infty} \simeq \delta$, which indicates that at finite doping and in the $U\rightarrow \infty$ limit only the {\it ferromagnetic} interaction survives. This %is potentially linked
links to the Nagaoka ferromagnetism\cite{Nagaoka1966,Tasaki1998}. % in the thermodynamic limit and at finite doping.
%This result may indicate that solving the HM dynamically then take the $U\rightarrow \infty$ limit is not equivalent to enforcing the Gutzwiller-projection directly in the thermodynamic limit, which .

\begin{figure}
  \centering
  %\subfigure[]{\label{subfig:J-colormap}\includegraphics[width=.45 \columnwidth]{./Jplot.png}}
  %\subfigure[]{\label{subfig:J-delta}
    \includegraphics[width=.6 \columnwidth]{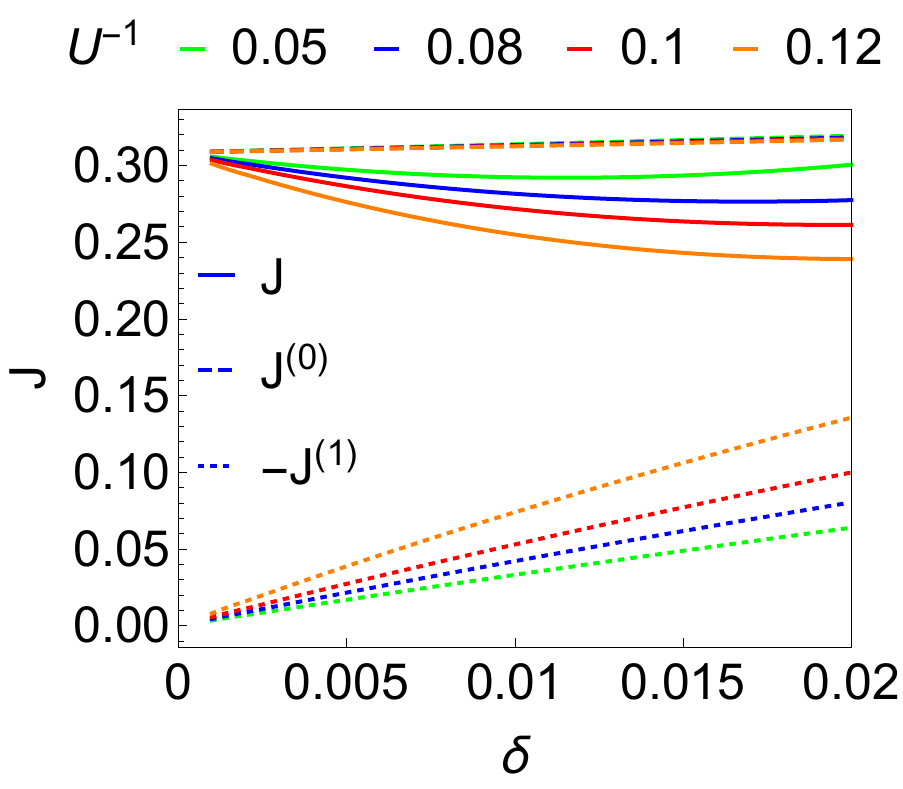}
  \caption{ (i) the total spin-spin interactions strength $J$ (solid lines), (ii) the superexchange interaction strength $J^{(0)}$ (dashed lines) and (iii) the ferromagnetic interaction strength $(-1) \times J^{(1)}$ (dotted lines) plotted as functions of $\delta$.}
  \label{fig:Jplot}
\end{figure}

{\it Effective theory at first order in $t/U$.~~}
%With the solution of the dynamics of the doped mean field theory,
Based on the above mean-field solution, we can further construct an effective theory %theories
for the doped HM in a way similar to %regime in the similar way as we proposed in
\refdisp{Ding2014}.  %and is only implemented through the computation of the physical electronic Green's functions.
%With the results in the previous sections, we propose to extend the saddle point theory given in \eqdisp{eq:H-S-eff-0-order} and (\ref{eq:H-f-eff-0-order}) to a set of second order effective theory which includes the dynamical fluctuations as the following:
The resulting theory contains Hamiltonians in both the slave-spin and $f$-spinon sectors
\begin{align}
  H_{S,eff} = H_{S,0} + H_{S,t} , \\
  H_{f,eff} = H_{f,t} + H_{f,0} + H_{J}.
\end{align}
$H_{f,eff}$ takes the form of a %conventional
\tJ model and $H_{S,eff}$ takes the form of an XXZ spin model. %-type slave-spin modelbut with dynamically renormalized $t$ and $J$.
They are dynamically coupled via self-consistent condition.
%We obtain a renormalized \tJ -like model for the $f$-spinons coupled %and
%to an XXZ-type slave-spin model. %theory, which are self-consistently coupled. Note that
In our theory the renormalization factors of $t$ and $J$ for
the $f$-spinons, $g_t$ and $g_J$, are via the slave-spin correlations and no further
%{\it not} Gutzwiller-projected. The
Gutzwiller projection is necessary. %automatically encoded in the slave spin correlations.
Quantitatively we find %our dynamical $t/U$ expansion theory gives
\begin{align}
  g_t = M_x^2 = 2 \delta, \quad g_J = 1- 2 \delta,
\end{align}
which agree well to those
%Comparing with the
phenomenological values $g_t = 2 \delta / (1 + \delta), \quad
  g_J= 1 / (1+ \delta)^2$ %renormalization of $t$ and $J$
first proposed by Zhang et. al.~\cite{Zhang1988}. %, in which $
%  g_t = 2 \delta / (1 + \delta), \quad
%  g_J= 1 / (1+ \delta)^2,
%$
%our dynamical $t/U$ expansion theory gives
%\begin{align}
%  g_t = M_x^2 = 2 \delta, \quad g_J = 1- 2 \delta.
%\end{align}
%where we absorb the factor $(1 - 4 a^2)^{-1}$ into the bare $J_0$ for $g_J$.
%$g_t$ is given by the static component of $S^{\pm}$, i.e. the quasiparticle weight $Z = M_x^2/2$. Both factors agree with and explains the Gutzwiller approximation proposed in \refdisp{Zhang1988} at the leading order in $\delta$.

{\it %Discussion
Conclusion.~~}
In this work, we propose a slave-spin representation of the HM on a square lattice with %only
nn hoppings and use the SEoM perturbation theory to implement a dynamical $t/U$ expansion in the doped MI. %treat the slave spin sector of the slave spin representation of HM on a square lattice with %only
%nn hoppings. Although we only solve the doping driven metallic phase at a previously known,
At the saddle-point level, our theory %level with a mean field treatment for the hopping term, we already obtain
generates nontrivial dynamical properties of single-electron spectrum, including multiple pole structures and spectral weight (re-)distributions %as
with doping, %varies,
Both features %of which
are in excellent agreement with %previous
known numerical results %. However,
despite that our theory is analytic in nature and works directly in the $\omega-\k$ space. \ry{We also derive the exchange interactions among spins. In addition to the usual AFM superexchange interactions, we find a new FM channel at finite doping and finite $U$ values. This FM coupling survives the $U\rightarrow \infty$ limit and asymptotically connects to the Nagaoka ferromagnetism. It also provides a viable explanation for recently discovered FM %phase
order\cite{Sarkar2019} and %FM-type
spin fluctuations\cite{Sonier2010,Butch2012,Kurashima2018} in %the
cuprates.
In general, we} find the low-energy effective theory of the doped MI is a dynamical \tJ model with effective renormalization factors in agreement with those % of the parameters for the effective theory, i.e. the effective $f$-spinon \tJ model, is in agreement with what is
proposed phenomenologically %by Zhang et. al.
based on experimental data and empirical findings. Our theory thus provides a natural basis and reliable means in understanding the exotic properties of the doped HM.

{\it Acknowledgement:} We thank Q. Si for motivating this work, and J. Wu for useful discussions. The work at Anhui University was supported by the Startup Grant number S020118002/002 of Anhui University. WD thanks support from Kavli Institute for Theoretical Sciences. The work at Renmin University was supported by the National Science Foundation of China Grant number 11674392, the Fundamental Research Funds for the Central Universities and the Research Funds of Remnin University of China Grant number 18XNLG24, and the Ministry of Science and Technology of China, National Program on Key Research Project Grant number 2016YFA0300504.
\newpage

\setcounter{equation}{0}
\setcounter{figure}{0}
\setcounter{table}{0}
\setcounter{page}{1}
\makeatletter
\renewcommand{\theequation}{S\arabic{equation}}
\renewcommand{\thefigure}{S\arabic{figure}}
\renewcommand{\bibnumfmt}[1]{[S#1]}
\renewcommand{\citenumfont}[1]{S#1}
\renewcommand \thetable{S\@arabic\c@table}
\newcommand{\sref}[1]{({\color{blue}S}\ref{#1})}
\newcommand{\sfigdisp}[1]{Fig.~({\color{blue}S}\ref{#1})}

\onecolumngrid
\begin{center}
  \textbf{\Large Supplemental Materials}
\end{center}

\vspace{0.5cm}

\twocolumngrid
%%%%%%% Widetext - title %%%%%%%%%%

%{\it Formal self-consistent decoupling of the slave spin and spinons:~~}

\section*{ %Weiss mean field theory results
Solution of the slave-spin theory within Weiss mean-field approximation
}
In this part, we show the \ry{saddle-point solution of the slave spin theory within a Weiss mean-field decomposition in Eq.~(7) of the main text} %Weiss mean field theory results
obtained by diagonalization.

\begin{figure}
  \centering
  \subfigure[]{\label{SM-subfig:h-delta-plot}\includegraphics[width=0.48 \columnwidth]{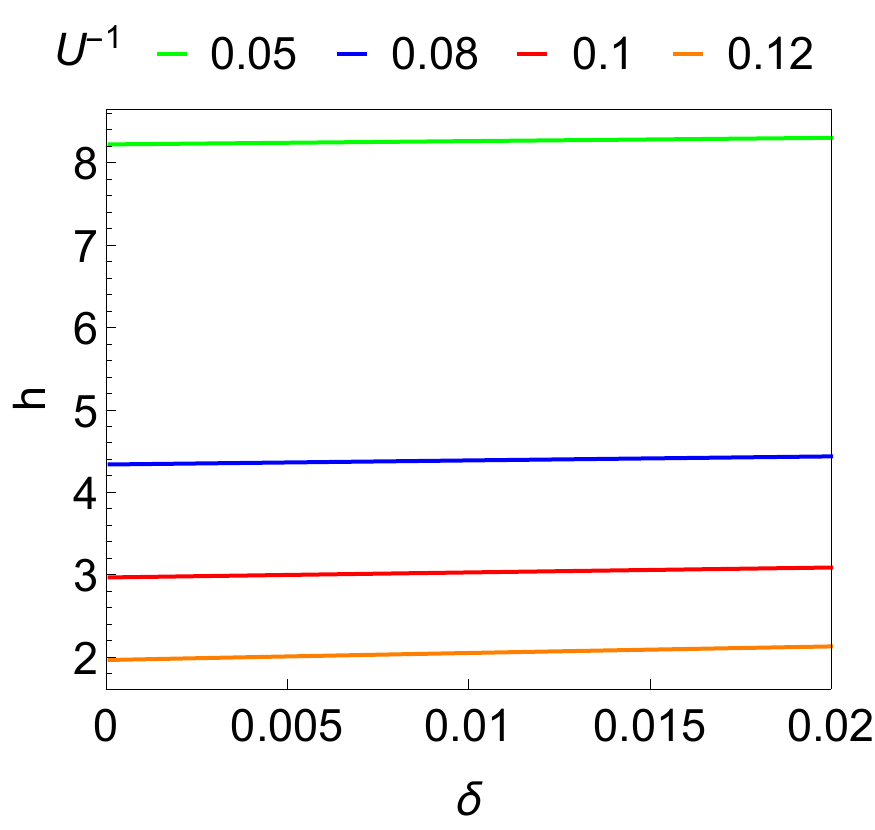}}
  \subfigure[]{\label{SM-subfig:hc-plot}\includegraphics[width=0.48 \columnwidth]{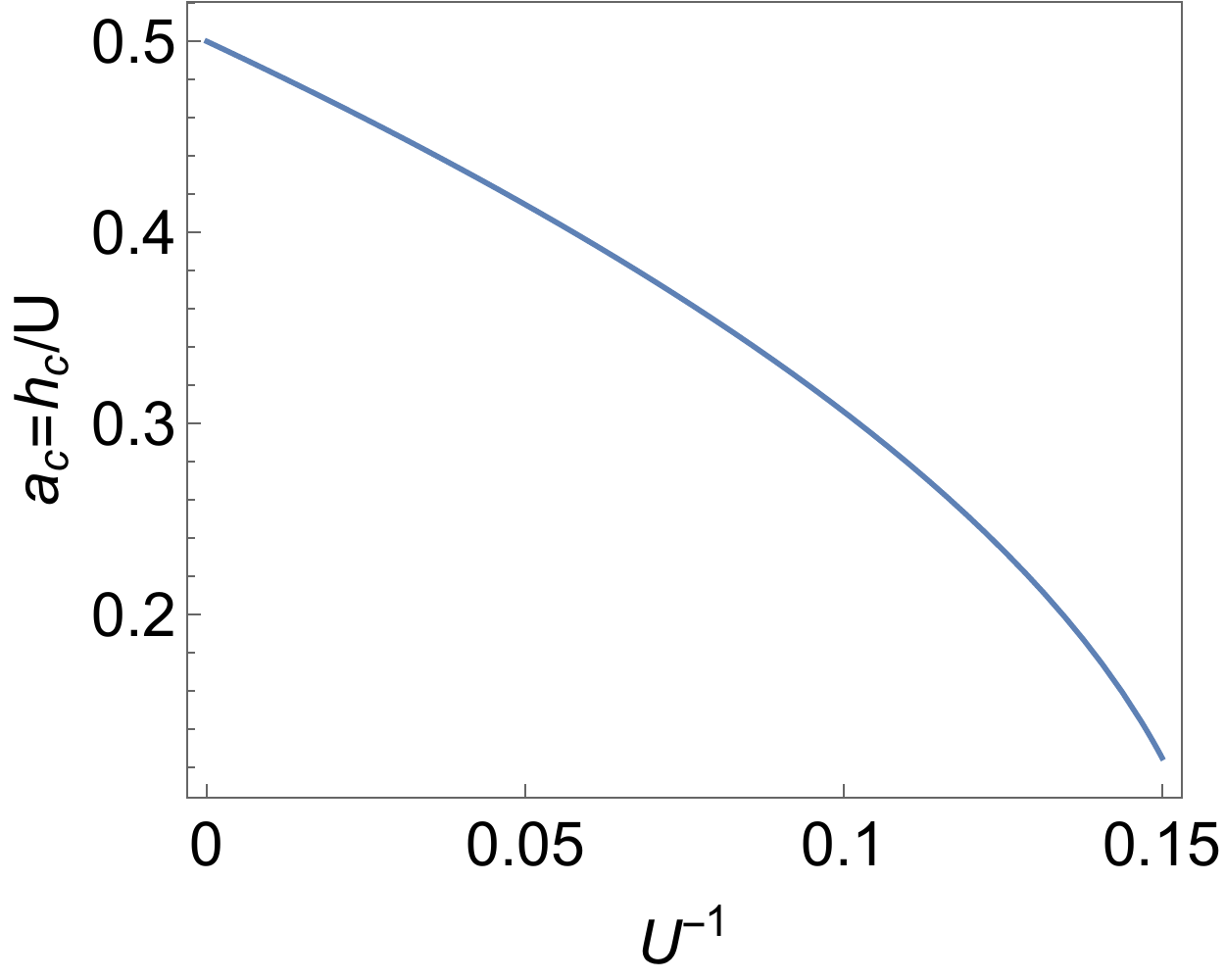}}
  \subfigure[]{\label{SM-subfig:ad-plot}\includegraphics[width=0.48 \columnwidth]{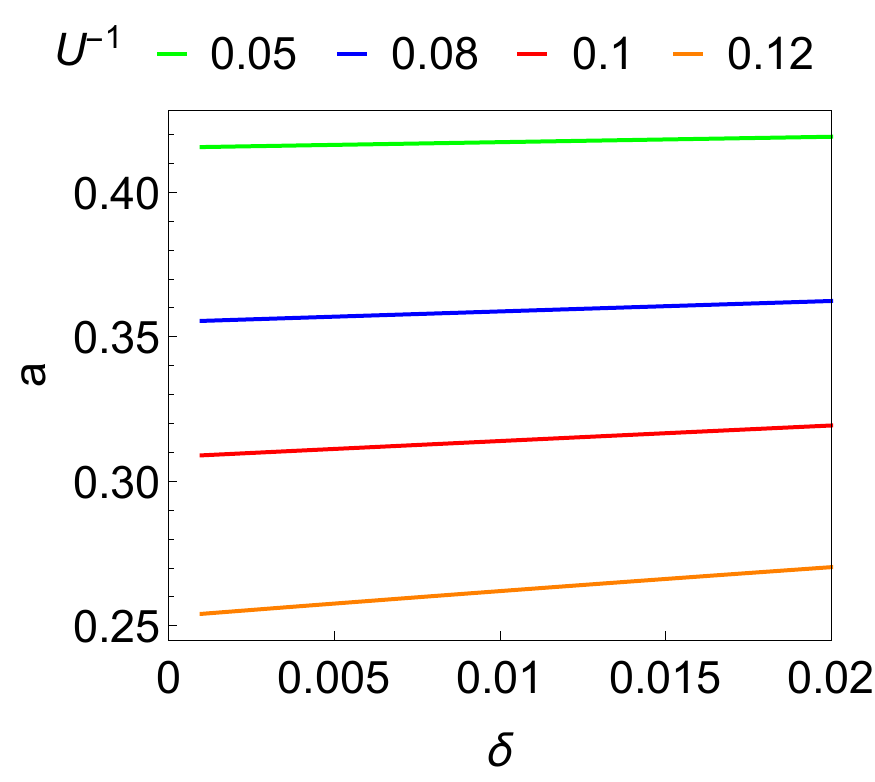}}
  \subfigure[]{\label{SM-subfig:delta-A-plot}\includegraphics[width=0.48 \columnwidth]{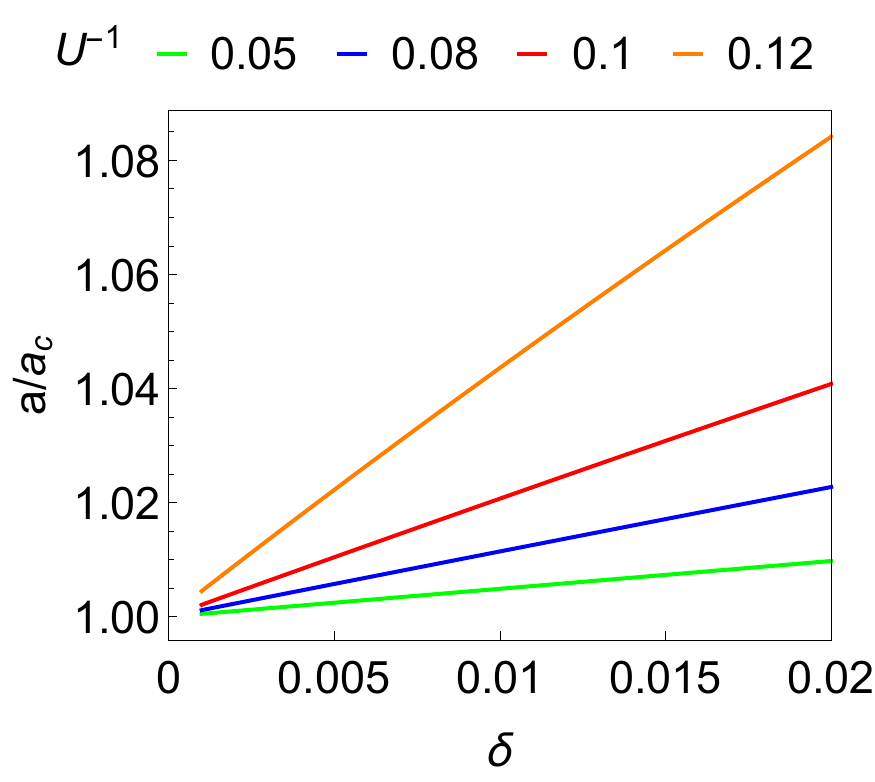}}
  \caption{\sref{SM-subfig:h-delta-plot} bare values of $h$ shown at different $U$'s as functions of $\delta$; \sref{SM-subfig:hc-plot} $a_c = h_c/U$ shown as a function of $U^{-1}$ over the full range; \sref{SM-subfig:ad-plot} bare values of $a = h/U$; \sref{SM-subfig:delta-A-plot} the relative change of $a$ from $a_c$}
\end{figure}

First, we find the dMIMT %happens
\ry{taking place}
at a finite $h_c$ with
\begin{align}
h_c \simeq U/2 \times \sqrt{U_c(U^{-1}_c - U^{-1})}.
\end{align}
Note that in the single-orbital Hubbard model, one can always set $h = - \mu$, where $\mu$ is the %physical
electron chemical potential. This makes the ratio $a=h/U$ proportional to doping $\delta$, as shown in Figs.~\sfigdisp{SM-subfig:h-delta-plot} and ~\sfigdisp{SM-subfig:hc-plot}. Actually,
\begin{align}
a = a_c (1 + b~\delta).
\end{align}
\ry{where the factor $b$ decreases as $U$ increases, and the critical value $a_c = \sqrt{U^{-1}_c - U^{-1}}/2$.}

The ratio \ry{$a$} %$a = h /U$
is shown for both its bare values in \sfigdisp{SM-subfig:ad-plot} and its changes from the critical point $a/a_c$ in \sfigdisp{SM-subfig:delta-A-plot}. The change in $a$ as a percentage of $a_c$ is about 4 times of $\delta$, hence leads to the increase of $J_0$ as $\delta$ increases. We consider that it is  an artifact of mean field theory since $h$ accounts also for effects due to the hopping terms.

%The dMIMT is a very weakly first order transition as our calculation for $\delta$ up to $ \delta = 10^{-5}$ gives $M_x \simeq 0.0027$ at $U=10$, similar to the previous 5(2)-orbital calculation?[refs]
%We show results for $M_x$ at several different $U$'s in \sfigdisp{SM-subfig:Mx-plot} and $ M_{x,C}\simeq M_{x}\vert_{\delta = 10^{-5}}$ in \sfigdisp{SM-subfig:Mxc-plot}.

%\begin{figure}
%  \centering
%   \subfigure[]{\label{SM-subfig:Mx-plot}\includegraphics[width=0.48 \columnwidth]{./Mxplot.pdf}}
%   \subfigure[]{\label{SM-subfig:Mxc-plot}\includegraphics[width=0.48 \columnwidth]{./Mxcplot.pdf}}
%  \caption{ \sref{SM-subfig:Mx-plot} $M_x$ shown at several different $U$'s, and \sref{SM-subfig:Mxc-plot} $M_{x,C}\simeq M_{x}\vert_{\delta = 10^{-5}}$. We find the dMIMT a very weakly first order transition.}
%\end{figure}
%

\begin{figure}
  \centering
 \subfigure[]{\label{SM-subfig:Z-plot}\includegraphics[width=0.48 \columnwidth]{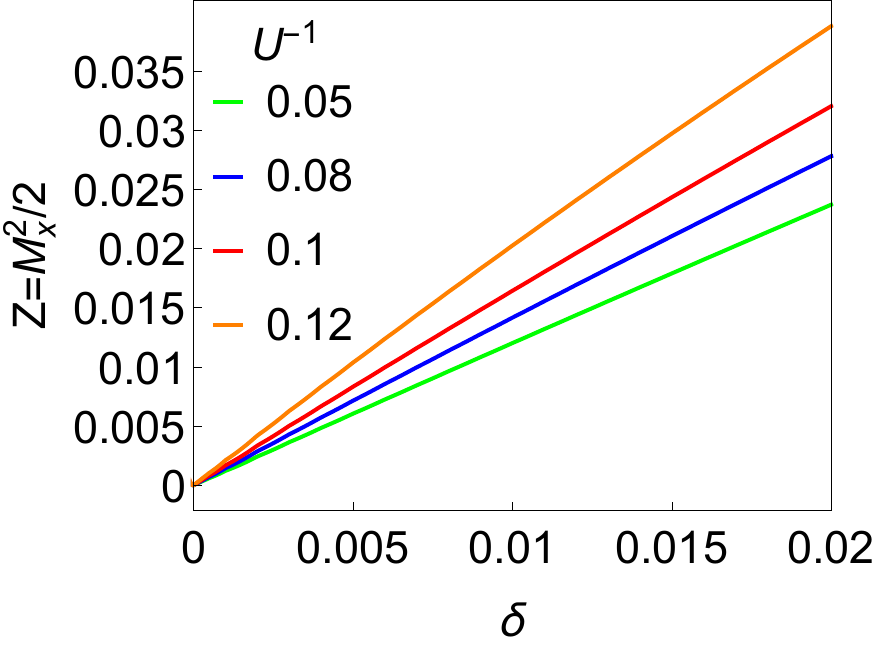}}
  \subfigure[]{\label{SM-subfig:ZU-plot}\includegraphics[width=0.48 \columnwidth]{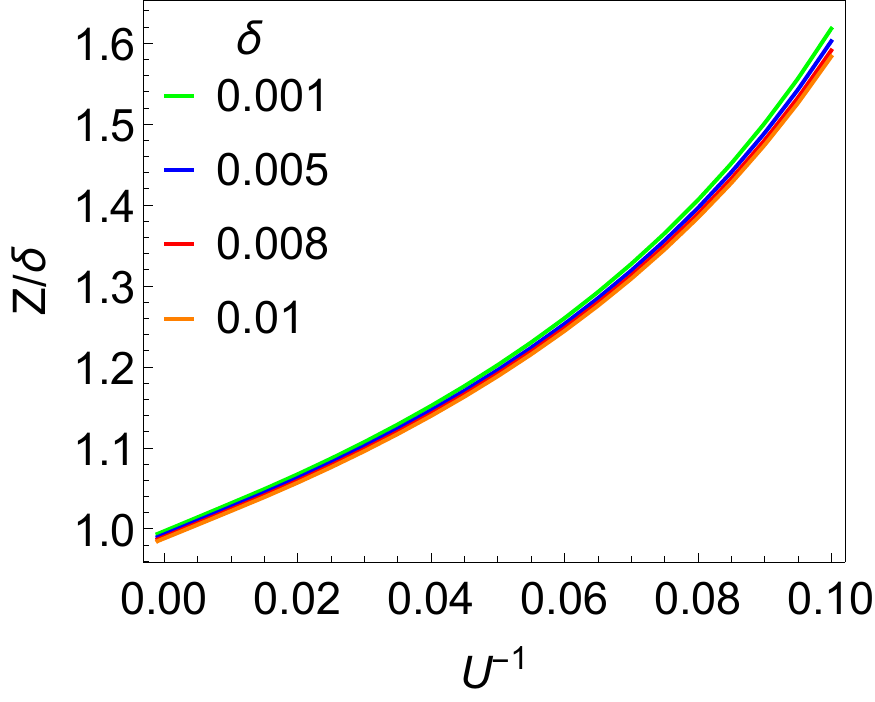}}
 %\subfigure[]{\label{SM-subfig:ZA-plot}\includegraphics[width=0.48 \columnwidth]{./firstOrderplot.pdf}}
  \caption{\sref{SM-subfig:Z-plot} $Z = M_x^2/2$ shown as functions of $\delta$ at different $U$'s; \sref{SM-subfig:ZU-plot} $Z/\delta$ plotted as a function of $U^{-1}$ down to zero which converges to $1$.}
\end{figure}

The dMIMT is \ry{of} Brinkman-Rice type, %mean field type transition,
as we find that $Z\propto \delta$ which is shown in \sfigdisp{SM-subfig:Z-plot}. To support the survival of the Nagaoka-ferromagnetic interaction in the $U\rightarrow \infty$ limit, we plot $Z/\delta$ as a function of $U^{-1}$. \ry{As shown in \sfigdisp{SM-subfig:ZU-plot}, $Z/\delta$
%down to zero which
converges to $1$ in the limit of $U\rightarrow\infty$, which indicates that the FM exchange coupling in Eq.(18) of the main text keeps finite in this limit.} %\sfigdisp{SM-subfig:ZU-plot}.

\section*{Perturbative Schwinger's equation-of-motion approach for quantum spins}
A full description of the perturbative Schwinger's equation-of-motion approach for spin-$1/2$ quantum spins is given in Ref.~\onlinecite{Ding2019b} independently. Here we briefly present the approach and the solutions for the slave spins.

The Schwinger's equation-of-motion theory converts the operator Heisenberg-equations-of-motion (HEoM) into equations of motion for the Green's functions. For the quantum spins that %satisfy
\ry{obey} the $SU(2)$ Lie-algebra, we introduce both a bosonic and a fermionic Green's functions as the follows:

\begin{widetext}
\begin{align}
  \begin{split}
     iG^{O O'}_{\eta}[i,f] = \dev{\hat{O}_i[t_i]  \hat{O}'_f[t_f]}_{\eta}  =&  \ev{\Torder_\pm \left[\hat{O}_i[t_i] \hat{O}'_f[t_f] \right]} - C_\eta\ev{\hat{O}_i} \ev{\hat{O}'_f}  \\
     =& ~ \ev{\theta(t_i-t_f) \hat{O}_i[t_i] \hat{O}'_f[t_f] + \eta \theta(t_f-t_i) \hat{O}'_f[t_f] \hat{O}_i[t_i]} - C_\eta\ev{\hat{O}_i} \ev{\hat{O}'_f},
   \end{split}
\end{align}
\end{widetext}
where $\eta = B,~F$ as subscripts while $\eta = \pm$ correspondingly in the equations and $C_{B(F)} = 2 (0)$. Whereas $G^{O O'}_{B(F)}[i,f]$ are considered not independent, we consider both here since sometimes it is more convenient to use one not the other for computing certain quantities.

\subsection{Atomic limit solution}
%For
\ry{In} the atomic limit, since the slave spin Hamiltonian
\begin{equation}
  \label{Seq:H-atom-1-orb}
  H_{S, at} = \frac{U}{2} \sum_i (\sum_{s} S_{is}^z)^2 + h \sum_{i s} S_{is}^z
\end{equation}
is purely Ising-type, we only need to consider
\begin{align}
G^{\alpha \bar{\alpha'}}_{\eta, S, s s'}[i,f] =\ev{\mathcal{T}[S_{i s}^\alpha (t_i) S_{f s'}^{\bar{\alpha}'} (t_f)]} - 2 \ev{S_{i s}^\alpha} \ev{S^{\bar{\alpha}'}_{f s'}},
\end{align}
with $\alpha = +$ or $-$.

First, we obtain the HEoM
\begin{align}
  -i\pt_t S^\alpha_{i s} = [H_{int}^S, S^\alpha_{i s}] = \alpha U S^z_{i \bar{s}} S^\alpha_{i s} + \alpha h S^\alpha_{i s}.
\label{Seq:atomic-Hs-HEoM}
\end{align}
Correspondingly, the SEoM is
\begin{widetext}
\begin{align}
   - i \pt_{t_i} G^{\alpha \bar{\alpha'}}_{B,S, \si \si'}[i,f] & = \alpha \left( 2 \ev{S^z_{\si}} \delta_{\alpha \alpha'} \delta_{\si \si'} \delta[i,f] - h G^{\alpha \bar{\alpha'}}_{B,S, \si \si'}[i,f]   + J ~\Gamma^{z \alpha; \bar{\alpha'} }_{B,S,\sib \si;\si'}[i,f] \right),\\
  % - i \pt_{t_i} G^{-+}_{B,S, \si \si'}[i,f] & = -2 \ev{S^z_{\si}} \delta[i,f] \delta_{\si \si'}  - h G^{-+}_{B,S, \si \si'}[i,f]  - J \Gamma^{z-;+}_{F,S,\sib \si;\si'}[i,f],\\
  - i \pt_{t_i} G^{\alpha \bar{\alpha'}  }_{F,S, \si \si'}[i,f] & = ~\delta_{\alpha \alpha' } \delta_{\si \si'} \delta[i,f]   + \alpha \left( - h G^{\alpha \bar{\alpha'} }_{F,S, \si \si'}[i,f]  + J ~\Gamma^{z\alpha; \bar{\alpha'}}_{F,S, \sib \si;\si'}[i,f] \right),
 % - i \pt_{t_i} G^{-+}_{F,S, \si \si'}[i,f] &=~ \delta[i,f] \delta_{\si \si'} - h G^{-+}_{F,S,\si \si'}[i,f] - J \Gamma^{z-;+}_{F,S,\sib \si;\si'} [i,f],
\end{align}
\end{widetext}
where $\Gamma^{\alpha \alpha' ;\alpha''}_{B(F) s s'; s''}[i,f]$ denotes the vertex functions defined as
\begin{align}
    i \Gamma^{\alpha \alpha' ;\alpha''}_{B(F),S, s s'; s''}[i,f] = \dev{S^\alpha_{s}[t_i] S^{\alpha'}_{s'}[t_i] S^{\alpha''}_{\s''}[t_f]}_{B(F)} .
\end{align}
In the Ising limit, the vertex function can be simplified as
\begin{align}
   i \Gamma^{z \alpha ;\alpha'}_{B(F),S,s s'; s''}[i,f] = \ev{S^z_{i s}} G^{\alpha \alpha'}_{B(F),S, s' s''}[i,f].
\end{align}

 To simplify the notation, we shall drop the slave spin index $s$ so that $G^{\alpha \bar{\alpha'}}_{S}$ indicates a $2\times 2$ matrix.
Here $\si_i$ denotes the Pauli matrices ($\si_0$ being the identity matrix). Denoting
$
  \ev{S^z_{a} + S^z_{b}}= M,\
  \ev{S^z_{a} - S^z_{b}}= m,
$ which are good quantum numbers,
%and note
%\begin{align}
%  m^2 + M^2 \le 1.
%\end{align}
%due to possible fluctuations.
we obtain
\begin{align}
   G^{\alpha \bar{\alpha'}}_{B,S}[\omega ] = \frac{\alpha \delta_{\alpha \alpha'}  (M  \si_0 + m  \si_z )}{\omega - \alpha ( h + U  (M  \si_0 - m  \si_z )/2)},\\
   G^{\alpha \bar{\alpha'}}_{F,S}[\omega ] = \frac{\delta_{\alpha \alpha'} \si_0}{\omega - \alpha ( + h + U  (M  \si_0 - m  \si_z )/2)}.
\end{align}
where inside the equation, $s = \pm 1$. \ry{What is $s$ here since $s$ has been just dropped?}

\subsubsection*{Expressions for arbitrary states }
For an arbitrary state $\ket{\psi}$, we can always expand it in terms of eigenstates of $H$. In this Ising limit, it is easy to prove that no crossing propagators for $\dev{S^+[t_i] S^-[t_f]}$ and $\dev{S^-[t_i] S^+[t_f]}$. Therefore, for any other states $\ket{\psi(M,~ m, ~\Delta m)}$ as given below,
  \begin{align}
    \ket{\psi_{M, m,  \Delta m}} = a \ket{\ua \da} + b \ket{\da \ua}+ c \ket{\ua \ua} + d \ket{\da \da},
  \end{align}
  where \begin{align}
          M = c^2 - d^2, \quad m = a^2 - b^2, \nonumber \\
          \Delta m = \sqrt{\ev{\hat{m}^2}} = \sqrt{a^2 + b^2}.
  \end{align}
We find that the arbitrary $G^{\alpha \alpha'}_{B(F), para}$ can \ry{be} constructed as
  \begin{equation}\label{eq:gen-G-construction}
    \begin{split}
      & G^{\alpha \alpha'}_{B(F), S, para} = \\
      & a_1^2 G_{B(F),S, M=1, m=0} + a_2^2 G_{B(F),S, M=-1, m=0} \\
      & + a_3^2 G_{B(F),S, M=0,m=1} + a_4^2 G_{B(F), S, M=0,m=-1},
  \end{split}
\end{equation}
where $para = (M, m,  \Delta m)$ is the \ry{complete parameter set that %completes specify 
describes} the underlying state.
With \eqdisp{eq:gen-G-construction}, we can plug the $G_{B(F),S}$ back into the SEoM to obtain solutions for the vertex functions.

To prepare for the perturbation calculation of transverse field, we write down the explicit expressions for arbitrary states with physical parametrization (use $(M,~m,~\Delta m)$ instead of $a_i$s).

First, the solution for real $a_i$s %are 
\ry{is} not unique. For later purpose, here we pick a solution that gives us a positive and uniform $\ev{S^x}$:
\begin{align}
  a_1 = \sqrt{\frac{1+M-\Delta m^2}{2}},\quad a_2 = \sqrt{\frac{1-M-\Delta m^2}{2}},\\
  a_3 = \sqrt{\frac{m + \Delta m^2}{2}}, \quad a_4 = \sqrt{\frac{\Delta m^2 - m}{2}},
\end{align}
which gives
\begin{align}
  \begin{split}
    & \ev{S^x_a} =  \ev{S^x_b} = 1/2 (\sqrt{\Delta m^2 + m} \sqrt{1 - \Delta m^2 - M}\\
    &+ \sqrt{\Delta m^2 - m} \sqrt{1 - \Delta m^2 + M}).
 \end{split}
\end{align}
Now we can write \eqdisp{eq:gen-G-construction} as
\begin{align}
  \begin{split}
    & G^{\alpha \bar{\alpha}'}_{B(F)} = \frac{1}{2} \Big((1-\Delta m^2) (G^{\alpha \bar{\alpha}'}_{B(F),S,(1,0)}  + G^{\alpha \bar{\alpha}'}_{B(F),S,(-1,0)}) \\
    & + M  (G^{\alpha \bar{\alpha}'}_{B(F),S,(1,0)} - G^{\alpha \bar{\alpha}'}_{B(F),S,(-1,0)}) \\
    & +  m (G^{\alpha \bar{\alpha}'}_{B(F),S,(0,1)} - G^{\alpha \bar{\alpha}'}_{B(F),S,(0,-1)}) \\
    &+ \Delta m^2  (G^{\alpha \bar{\alpha}'}_{B(F),S,(0, 1)} + G^{\alpha \bar{\alpha}'}_{B(F),S,(0,-1)})\Big),
\end{split}
\end{align}
where $(M,m)$ denotes the base states parameters.

The $G^{\alpha \alpha'}_{B(F)}$ for arbitrary state with parameters $(M,\ m,\ \Delta m)$ read
\begin{widetext}
\begin{align}
  G^{\alpha \bar{\alpha}}_{B,S,0}[\omega] = \frac{\alpha ( (1 - \Delta m^2) \si_0 + m \si_z) (\omega + \alpha h) + (M + \alpha \Delta m^2) J \si_0 /2}{(\omega + \alpha h)^2 - U^2/4}, \label{eq:G_B0}\\
  G^{\alpha \bar{\alpha}}_{F,S,0}[\omega] = \frac{ (\omega + \alpha h) \si_0 + \alpha (M \si_0 -  m \si_z) J /2}{(\omega + \alpha h)^2 - U^2/4}. \label{eq:G_F0}
\end{align}
\end{widetext}

%%%%%%%%%%%%%%%%%%%%%%%%
\subsection*{Weiss mean-field %theory
approximation to %of
the hopping term of %for
the slave spin Hamiltonian}
In this part, we solve $H_{S,eff}$ for finite doping  at the mean-field level. The mean-field approximation is to decouple $H_{S,hopping}$ as
   \begin{align}
     \begin{split}
       &H_{S,hopping} \rightarrow H_{S,hMF} \\
       & = - Q_f \sum_{is} \big( (\sum_{\ev{ij} s'} \ev{S^-_{js'}}) S^+_{is} + h.c. \big),
  \end{split}
   \end{align}
   Since the emergence of $\ev{S_{i s}^{\pm}} \neq 0$ is from spontaneous-symmetry-breaking, we can choose the direction of the magnetization at our convenience: $\ev{S^+_{is}} = \ev{S^-_{is}} = \ev{S^x_{is}} = M_x$. On a 2D square lattice with only nearest neighbor (nn) hopping, $H_{S,MF}$ becomes
\begin{align}
  \begin{split}
 & H_{S,MF} = H_{S,at} + H_{S,hMF}\\
 & = \sum_{i} \Big( U S^z_{ia} S^z_{ib} + h (S^z_{ia} + S^z_{ib} ) - h_x (S^x_{ia} + S^x_{ib}) \Big),
\end{split}
\end{align}
where $h_x = D M_x Q_f$, $D=4$ is the number of nn bonds for a two dimensional square lattice. This is a local Hamiltonian and can be %easily
solved by exact diagonalization. %However, we shall our
Here we adopt an alternative analytic calculation using a perturbation theory in terms of the Green's functions.

The corresponding HEoM reads
\begin{align}
  -i\pt_t S^\alpha_{s} = \alpha (U S^z_{\bar{s}} S^\alpha_{s} + h S^\alpha_{s} + h_x S^z_{s} ),
  \label{eq:atomic-Hs-HEoM}
\end{align}
and hence the SEoM becomes
\begin{widetext}
\begin{align}
   - i \pt_{t_i} G^{\alpha \bar{\alpha'}}_{B,S, s  s '}[i,f] & = \alpha \left( 2 \ev{S^z_{s }} \delta_{\alpha \alpha'} \delta_{s  s '} \delta[i,f] - h G^{\alpha \bar{\alpha'}}_{B,S, s  s '}[i,f]   + J~ \Gamma^{z \alpha; \bar{\alpha'} }_{B,S,\bar{s} s ;s '}[i,f] + h_x G^{z \bar{\alpha'}}_{B,S,\bar{s} s'}[i,f] \right), \label{eq:transverse-field-B-EOM}\\
  - i \pt_{t_i} G^{\alpha \bar{\alpha'}  }_{F,S, s  s '}[i,f] & = ~\delta_{\alpha \alpha' } \delta_{s  s '} \delta[i,f]   + \alpha \left( - h G^{\alpha \bar{\alpha'} }_{F,S, s  s '}[i,f]  + J~ \Gamma^{z\alpha; \bar{\alpha'}}_{F,S, \bar{s} s ;s '}[i,f] +h_x G^{z \bar{\alpha'}}_{F,S,\bar{s} s'}[i,f] \right). \label{eq:transverse-field-F-EOM}
\end{align}
\end{widetext}
%\subsection{perturbation on product states}
According to the result of diagonalization, %calculation,  
we know that the ground state %to be the 
is a singlet/triplet with the onset of an infinitesimal transverse field.

The effects of the perturbation term are twofold: i) modifying the ground state wavefunction(s); ii) altering the evolution of the states (altering the SEoM). So we first consider the change in wavefunction, which gives $G_0$ with renormalized parameters. Then we consider the revised SEoM hence the further correction to $G$'s and $\Gamma$'s.

In the presence of a transverse field $h_x \hat{x}$, a magnetization $M_{x,s}$ along the field direction is induced. Now with the new correlators $G^{z \bar{\alpha'}}_{B(F)}$ entering the SEoM, we need to consider their HEoM and SEoM as well. The HEoM of $S^z_s $ reads
\begin{align}
  - i \pt_t S^z_s  = i h_x S^y_s  =  \frac{h_x}{2} (S^+_s  - S^-_{s }),
\end{align}
which leads to the following SEoM in frequency space
\begin{align}
 \omega G^{z \bar{\alpha'}}_{B}[\omega] &=  \frac{1}{2}(\alpha' I_{x} + h_x \sum_\alpha \alpha G^{\alpha \bar{\alpha'}}_B[\omega]), \label{eq:heom-G-B-za} \\
  \omega G^{z \bar{\alpha'}}_{F}[\omega] & =  \frac{h_x}{2} \sum_{\alpha} \alpha G^{\alpha \bar{\alpha'}}_F[\omega], \label{eq:heom-G-F-za}
\end{align}
where the second equal sign is because we apply a uniform field along $\hat{x}$.

Note that
\begin{align}
%  &\ev{S^z_{s}}  = \frac{1}{2} G_F^{+-}[i,i],\\
 \begin{split}
    \ev{S^x_s} & = -i \ev{[S^z_s, S^y_s]}  = \frac{1}{2 i} (G_F^{z+}[i,i] - G_F^{z-}[i,i])\\
    =& -\int \frac{d\omega}{2 \pi} \frac{h_x}{2 i \omega} (G_{F,S,s s}^{++}[\omega] + G_{F,S,s s}^{--}[\omega] \\
   &- G_{F,S,s s}^{-+}[\omega]  - G_{F,S,s s}^{+-}[\omega]).
 \end{split}
%  & \ev{S^z_{s} S^z_{sb}} - \ev{S^z_{s}} \ev{S^z_{sb}} = \int dt ~G^{+-}_{F,S,s s}[t] G^{+-}_{F,S, sb sb}[-t].
\end{align}
To the lowest order, the latter two terms can be computed as
\begin{align}
  \ev{S^x_s} = \int \frac{d\omega}{2 \pi} \frac{h_x}{2 i \omega} (  G_{F,S,0,s s}^{-+}[\omega] + G_{F,S,0,s s}^{+-}[\omega]), \label{eq:Sx-expr}
\end{align}
where $G_{F,S,0,s s}^{-+}[\omega]$ is the Green's function without transverse field in \eqdisp{eq:G_F0}.

The transverse magnetization, i.e. the quasiparticle weight, the magnetization, i.e. the hole density, and $\Delta m^2$ correction, to the lowest order in $h_x$, are found to be
\begin{align}
  \ev{S^x_{a}} \simeq h_x \left[\frac{U \si_0 - 2 h M \si_0 + 2 h m \si_z}{U^2 - 4 h^2}\right]_{aa},\\
  M = 2 \ev{S^z_a} \simeq \frac{- 4 h h_x^2}{U (U^2/4 - h^2)},\\
  \delta \Delta m^2 \simeq \frac{- h_x^2}{U^2/4 - h^2}.
\end{align}
which leads to $Z  = M_x^2/2 \propto \delta$ by solving the self-consistency equation $h_x = D M_x Q_f$. All are consistent with the numerical calculations.

%With 
In our SEoM theory, %we have expressions for 
the dynamical spin Green's functions of $H_{S,MF}$ can be written as %the following.
\begin{align}
  G^{\alpha \bar{\alpha'}}_{B(F)}[\omega] \simeq G^{\alpha \bar{\alpha'}}_{B(F),S,0}[\omega] + G^{\alpha \bar{\alpha'}}_{B(F),S,1}[\omega],
\end{align}
where $G^{\alpha \bar{\alpha'}}_{B(F),S,1}[\omega]$ is the lowest order correction of $G^{\alpha \bar{\alpha'}}_{B(F)}[\omega]$ (other than the change in the wavefunction under the evolution of $H_0$).
\begin{widetext}
\begin{align}
  \begin{split}
    & G^{\alpha \bar{\alpha'}}_{B,S,1, s s'}[\omega] = \frac{\alpha}{\omega + \alpha h} \Big( \frac{\alpha' (U \ev{S^x_{s}} + h_x) I_x}{2}  + \frac{U h_x}{2 \omega}  \ev{S^x_{s}}  \sum_{\alpha''} \alpha'' G^{\alpha'' \bar{\alpha'}}_{B,S,0, \bar{s} s'}[\omega]  + \frac{h_x^2}{2 \omega} \sum_{\alpha''} \alpha'' G^{\alpha'' \bar{\alpha'}}_{B,S,0, s s'}[\omega]\Big),
  \end{split}
\end{align}
\begin{align}
   \begin{split}
     & G^{\alpha \bar{\alpha'}}_{F,S,1, s s'}[\omega] = \frac{\alpha}{\omega + \alpha h} \Big( \frac{U h_x}{2 \omega} M_x  \sum_{\alpha''} \alpha'' G^{\alpha'' \bar{\alpha'}}_{F,S,0, \bar{s} s'}[\omega]  + \frac{h_x^2}{2 \omega} \sum_{\alpha''} \alpha'' G^{\alpha'' \bar{\alpha'}}_{F,S,0, s s'}[\omega]\Big).
\end{split}
\end{align}
%\begin{align}
%  G^{\alpha \bar{\alpha}}_{B,S,0}[\omega] = \frac{\alpha ( (1 - \Delta m^2) s_0 + m \si_z) (\omega + \alpha h) + (M + \alpha \Delta m^2) U \si_0 /2}{(\omega + \alpha h)^2 - U^2/4}, \label{eq:G_B0}\\
%  G^{\alpha \bar{\alpha}}_{F,S,0}[\omega] = \frac{ (\omega + \alpha h) \si_0 + \alpha (M \si_0 -  m \si_z) U /2}{(\omega + \alpha h)^2 - U^2/4}. \label{eq:G_F0}
%\end{align}
\end{widetext}
\end{document}